\documentclass[fleqn,usenatbib]{mnras}

\usepackage{newtxtext,newtxmath}

\usepackage[T1]{fontenc}

\DeclareRobustCommand{\VAN}[3]{#2}
\let\VANthebibliography\thebibliography
\def\thebibliography{\DeclareRobustCommand{\VAN}[3]{##3}\VANthebibliography}

\usepackage{graphicx}	
\usepackage{amsmath}	
\usepackage{amssymb}	

\newcommand{\tar}{J1110+4817}

\title[J1110+4817 as a CSO]{J1110+4817 -- a compact symmetric object candidate revisited}

\author[Krezinger et al.]{
M\'{a}t\'{e} Krezinger,$^{1,2}$
S\'{a}ndor Frey,$^{2,3}$\thanks{E-mail: frey.sandor@csfk.mta.hu} 
Tao An,$^{4}$\thanks{E-mail: antao@shao.ac.cn}
Sumit Jaiswal$^{4}$ and 
Yingkang Zhang$^{4,5}$
\\
$^{1}$Department of Astronomy, E\"{o}tv\"{o}s Lor\'{a}nd University, P\'{a}zm\'{a}ny P\'{e}ter s\'{e}t\'{a}ny 1/A, H-1117 Budapest, Hungary \\
$^{2}$Konkoly Observatory, Research Centre for Astronomy and Earth Sciences, Konkoly Thege Mikl\'{o}s \'{u}t 15-17, H-1121 Budapest, Hungary \\
$^{3}$Institute of Physics, ELTE E\"{o}tv\"{o}s Lor\'{a}nd University, P\'{a}zm\'{a}ny P\'{e}ter s\'{e}t\'{a}ny 1/A, H-1117 Budapest, Hungary \\
$^{4}$Shanghai Astronomical Observatory, Key Laboratory of Radio Astronomy, CAS, 80 Nandan Road, Shanghai 200030, China \\
$^{5}$University of Chinese Academy of Sciences, 19A Yuquanlu, Beijing 100049, China \\
}

\date{Accepted 2020 June 08. Received 2020 May 29; in original form 2020 April 27}

\pubyear{2020}

\begin{document}
\label{firstpage}
\pagerange{\pageref{firstpage}--\pageref{lastpage}}
\maketitle

\begin{abstract}
Compact symmetric objects (CSOs) are radio-emitting active galactic nuclei (AGNs) typically with a double-lobed radio structure confined to within 1~kpc. CSOs represent the earliest evolutionary phase of jetted AGNs. Some of them may eventually evolve into large-scale extended double sources, while others stall within the host galaxy and die out, depending on the longevity of nuclear activity, the jet power, and parameters of the surrounding galactic environment. Studying CSOs is a useful tool for understanding the evolution of the galaxies and the interactions between the jets and the medium of the host galaxy. Based on milliarcsec-resolution imaging observations using very long baseline interferometry (VLBI), it is not always straightforward to distinguish between a compact double-lobed or a core--jet structure. The quasar J1110+4817 was considered a CSO candidate in the literature earlier, but because of the lack of clear evidence, it could not be securely classified as a CSO. Here we present a comprehensive analysis of archival multi-frequency VLBI observations combined with accurate {\em Gaia} optical astrometric information. Lower-frequency VLBI images reveal an extended radio feature nearly perpendicular to the main structural axis of the source, apparently emanating from the brighter northern feature, that is rare among the known CSOs. While the presence of a binary AGN system cannot be fully excluded, the most plausible explanation is that J1110+4817 is a CSO.
\end{abstract}




\section{Introduction}
\label{intro}

Compact symmetric objects (CSOs) are powerful active galactic nuclei (AGNs) that are characterised with double compact lobes/hot spots with a small overall size less than $1$~kpc and inverted radio spectrum \citep{1994ApJ...432L..87W, 2012ApJ...760...77A}. 
Imaging the radio structure of CSOs need milliarcsec (mas) scale resolution that can only be achieved by the very long baseline interferometry (VLBI) technique. 
The core, where the powerful jets are launched from, is usually weak or even invisible in VLBI images due to the strong synchrotron self-absorption or free--free absorption, or a mixture of these two \citep{2003AJ....126..723T,2006A&A...450..959O}. 
The radio emission mainly comes from hot spots where the jet head interacts with the surrounding interstellar medium (ISM). In terms of evolution, CSOs, depending on their radio power, could be the progenitors of Fanaroff--Riley type I or II \citep{1974MNRAS.167P..31F} radio galaxies \citep[e.g.][]{1996ApJ...460..634R, 2012ApJ...760...77A}. While they are compact sources, most CSOs show very little relativistic beaming effect suggesting their jet orientation is close to the plane of the sky. Kinematic studies \citep{1996ApJ...460..612R,1998A&A...337...69O,2005ApJ...622..136G,2009AN....330..149P,2012ApJS..198....5A} proved that these sources are young, only up to a few thousand years old, making CSOs excellent targets to study the early-stage evolution of extragalactic sources and the properties of the ISM in their host galaxies. Measurements \citep[e.g.][]{2003PASA...20...69P, 2012ApJS..198....5A} show the hot spots having advance speeds from $\mu$as\,yr$^{-1}$ to mas\,yr$^{-1}$ scale, although it depends on the inclination of the structure with respect to the line of sight and the redshift of the source. 

A ``CSO-like'' radio morphology could occasionally be observed in other types of AGNs. For example, a radio core and another component in the jet (perhaps a bright standing shock, or an emission feature along the curving path enhanced by relativistic beaming at a section with a small inclination) could in principle produce a pair of VLBI components with similar brightness and slow relative proper motion \citep{1994ApJ...425..568C}. Interestingly, close-separation ($< 10$~pc) supermassive black hole binaries (SMBHBs), e.g., 0402+379 \citep{2006ApJ...646...49R}, can also show similar radio structure and inverted spectrum, suggesting that the radio activity of the SMBHBs is recently triggered by a galaxy merger \citep{2018RaSc...53.1211A}. The identification of the $\sim140$-pc separation binary within the alleged triple AGN system J1502+1115 \citep{2014Natur.511...57D} has been debated by \citet{2014ApJ...792L...8W} who interpreted the two compact flat-spectrum radio components as double hot spots in symmetric CSO lobes (i.e., fueled by a single central SMBH) based on an edge-brightened radio structure. In this scenario, the flat radio spectrum could be caused by free--free absorption in the ionised gas in the foreground. To tell genuine CSOs apart from other types of sources with similar morphology, we define CSOs as VLBI-detected compact AGNs with \textit{(i)} lobe-dominated radio structure associated with a single AGN and \textit{(ii)} an overall (projected) extent smaller than 1~kpc. 

A handful of VLBI surveys were conducted on CSOs. The largest one is the COINS (CSOs Observed in the Northern Sky) survey \citep{2000ApJ...534...90P} where multi-frequency polarimetric measurements were taken by the Very Long Baseline Array (VLBA) to distinguish between CSOs and core--jet sources. The COINS sample was further investigated by \citet{2005ApJ...622..136G}. Nearly 90 confirmed CSOs were also studied by \citet{2006MNRAS.368.1411A}, comparing them to the so-called medium-size symmetric objects (MSOs) extending to larger ($> 1$~kpc) scales. \citet{2016MNRAS.459..820T} constructed another sample from the VLBA Imaging and Polarization Survey \citep[VIPS, ][]{2007ApJS..171...61H}, which confirmed the presence of more CSO candidates and also detected polarization in two bright lobed sources, doubling the number of known polarized CSO cases. Detailed multi-frequency observations of individual CSOs can also provide useful knowledge about their nature \citep[e.g.][]{2013A&A...550A.113W,2014A&A...569A..22M, 2016AN....337...42R, 2020A&A...635A.185P}.

The quasar J1110+4817 was catalogued as a radio source already in the 1960s \citep[named as 5C~2.183,][]{1968MNRAS.139..529P}. Its optical counterpart, SDSS J111036.32+481752.3, is listed with a spectroscopic redshift $z^{\prime} = 6.168$ in the Sloan Digital Sky Survey (SDSS) Data Release 13\footnote{\url{http://skyserver.sdss.org/dr13/en/tools/explore/Summary.aspx?id=1237658612517307020}} \citep[DR13,][]{2017ApJS..233...25A}. However, this value is certainly a result of false automatic identification of emission lines. Here we adopt the considerably lower redshift value $z = 0.74$ measured by \citet{1996MNRAS.282.1274H} and \citet{2016MNRAS.462.1603Y} independently.

The FIRST (Faint Images of the Radio Sky at Twenty-one centimeters) survey \citep{1995ApJ...450..559B} image of J1110+4817 observed with the Very Large Array (VLA) at 1.4~GHz shows an unresolved radio structure on $\sim 5\arcsec$ scale with a flux density of $\sim$0.5~Jy. 
The compact radio source was imaged with the VLBA at several frequencies, e.g. 2.3~GHz \citep{2002ApJS..141...13B}, 4.8~GHz \citep{2007ApJ...658..203H,2016MNRAS.459..820T}, and 8.4~GHz \citep{2000ApJ...534...90P,2002ApJS..141...13B,2016MNRAS.459..820T}. In these images, its nearly symmetric structure is extended to $\sim 30$ milliarcsec (mas) in the northeast--southwest direction, with the northeastern feature being the brighter. The two dominant components appear connected with a weak jet-like emission. \citet{2000ApJ...534...90P} interpreted the VLBI structure as a core--jet, and no longer considered J1110+4817 as a CSO candidate. However, later \citet{2016MNRAS.459..820T} revisited the source as a possible CSO. Based on their new multi-frequency VLBI observations, they were unable to either confirm or refute that J1110+4817 is a CSO. The uncertainty was caused by the fact that the source was not detected in the observation at 15~GHz that could have pinpointed the location of the possible flat-spectrum core emission. 
The lack of the unambiguous identification of the core left J1110+4817 as an unconfirmed CSO candidate. 

In this paper, we present a detailed analysis of published and unpublished archival multi-frequency VLBI data of J1110+4817 (Sect.~\ref{vlbi}). We investigate the properties of the source (Sect.~\ref{properties}) and present different scenarios to explain the observations available to date (Sect.~\ref{cso}). Based on the AGN's accurate optical position, we argue that the object is most likely a CSO. We discuss this scenario in more details, and the possible role of {\em Gaia} optical astrometry in assessing the CSO nature of other candidate compact radio sources in Sect.~\ref{discussion}. Finally, we give a summary in Sect.~\ref{summary}.
We assume a standard flat $\Lambda$ Cold Dark Matter cosmology with $\Omega_{\rm{m}} = 0.3$, $\Omega_{\Lambda} = 0.7$, and $H_{\rm{0}} = 70$~km\,s$^{-1}$\,Mpc$^{-1}$. In this model, 1~mas angular size corresponds to 7.3~pc projected linear size at $z=0.74$, and the angular size distance is $D_\mathrm{A} = 1505.9$~Mpc \citep{2006PASP..118.1711W}.

\section{Archival VLBI observations}
\label{vlbi}

The data archive\footnote{\url{https://archive.nrao.edu}} of the U.S. National Radio Astronomy Observatory (NRAO) contains several VLBI data sets of J1110+4817 observed with the VLBA since 1996. They correspond to either published studies of the source \citep{2000ApJ...534...90P,2002ApJS..141...13B,2007ApJ...658..203H,2016MNRAS.459..820T} or unpublished projects. Among the latter, there are experiments where J1110+4817 was scheduled as a calibrator source and therefore was not in the focus of the given project. In addition, calibrated VLBA imaging data are available in the Astrogeo database\footnote{\url{http://astrogeo.org/cgi-bin/imdb_get_source.csh?source=J1110\%2B4817}}. These were obtained at different frequencies and epochs from 1996 to 2018.

For the purposes of this study, we selected, downloaded and analysed VLBA data from the NRAO archive from the years 2004 and 2005. These were chosen to complement published and already available VLBI data of J1110+4817 in terms of time span and frequency coverage. Our goal was to study possible long-term changes in the source structure and to obtain more information on its spectral properties.

The VLBA project BC120 (PI: S. Chatterjee) was observed on 2004 May 31 at 1.5~GHz frequency. The quasar J1110+4817 was used as a phase-reference calibrator for the pulsar B1112+50 as part of a series of experiments to precisely determine parallaxes and proper motions of pulsars \citep{2009ApJ...698..250C}. Several short ($\sim 1.5$~min) scans were spent on J1110+4817 during a 2-h time interval, with a total accumulated on-source time $\sim 40$~min. The observations were made with eight 8-MHz wide intermediate frequency channels (IFs) in right circular polarization, resulting in a total bandwidth of 64~MHz. All the ten 25-m diameter antennas of the VLBA participated in the experiment. 

\begin{figure*}
\centering
\includegraphics[width=0.45\textwidth, angle=0]{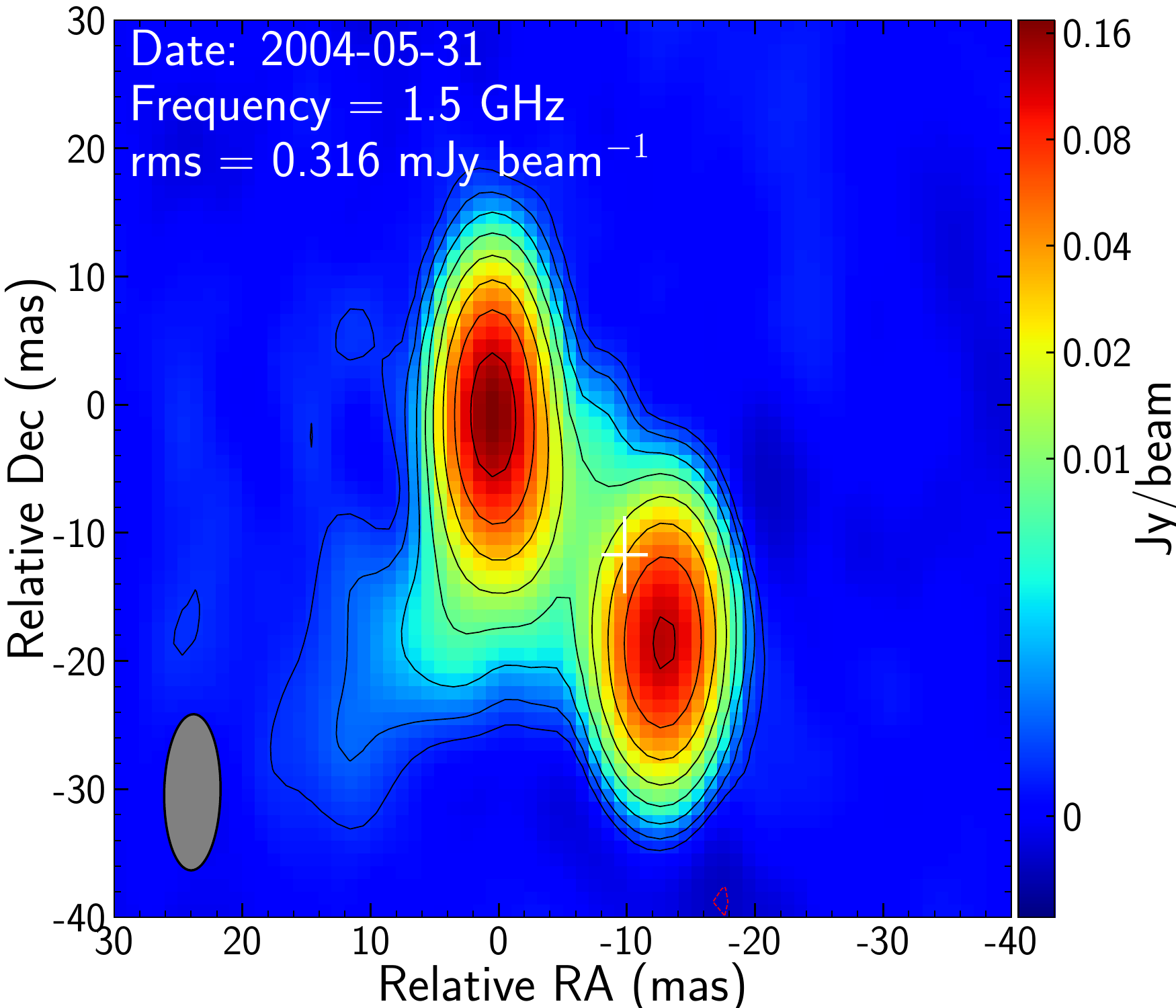} \hspace{5mm}
\includegraphics[width=0.45\textwidth, angle=0]{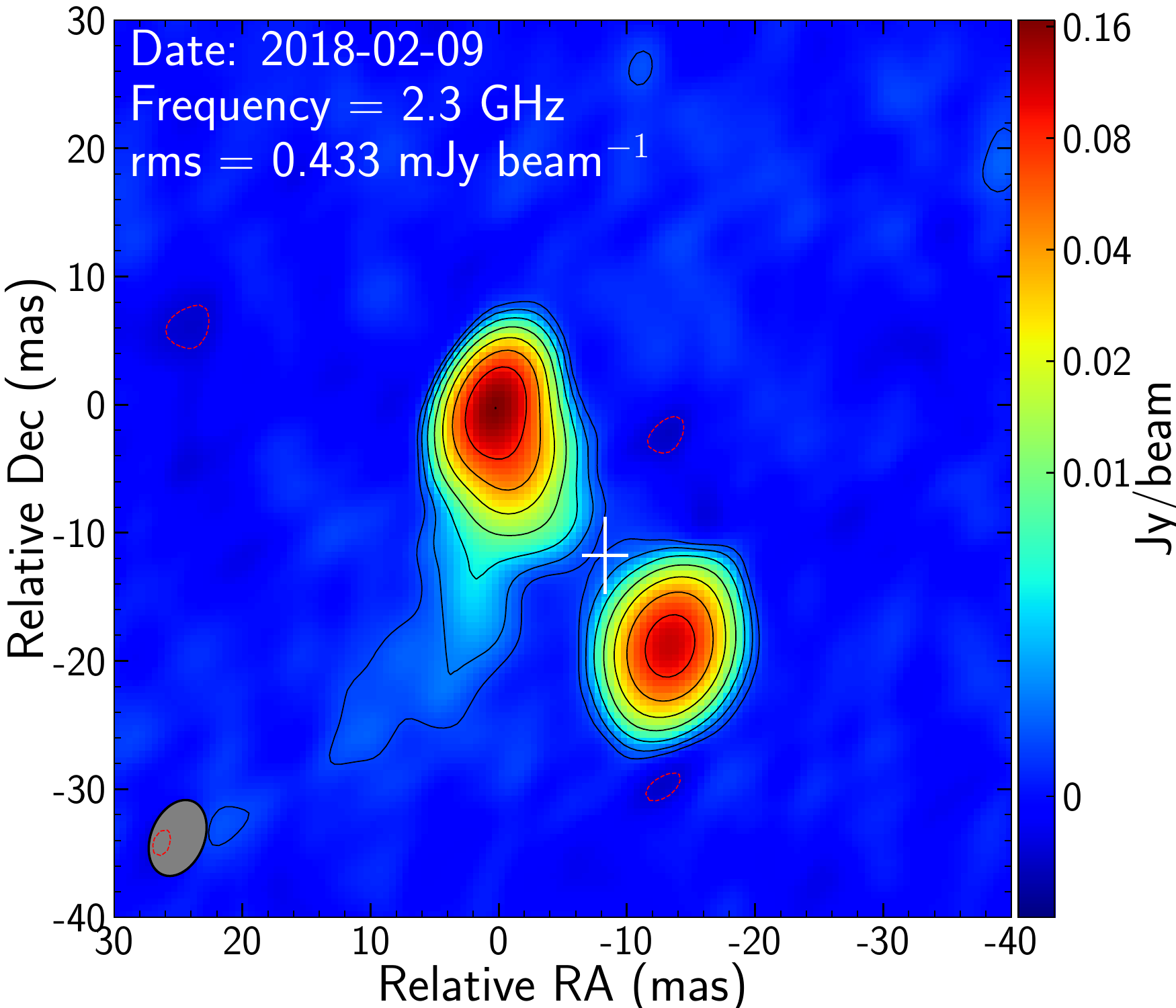} \\ 
\vspace{2mm}
\includegraphics[width=0.45\textwidth, angle=0]{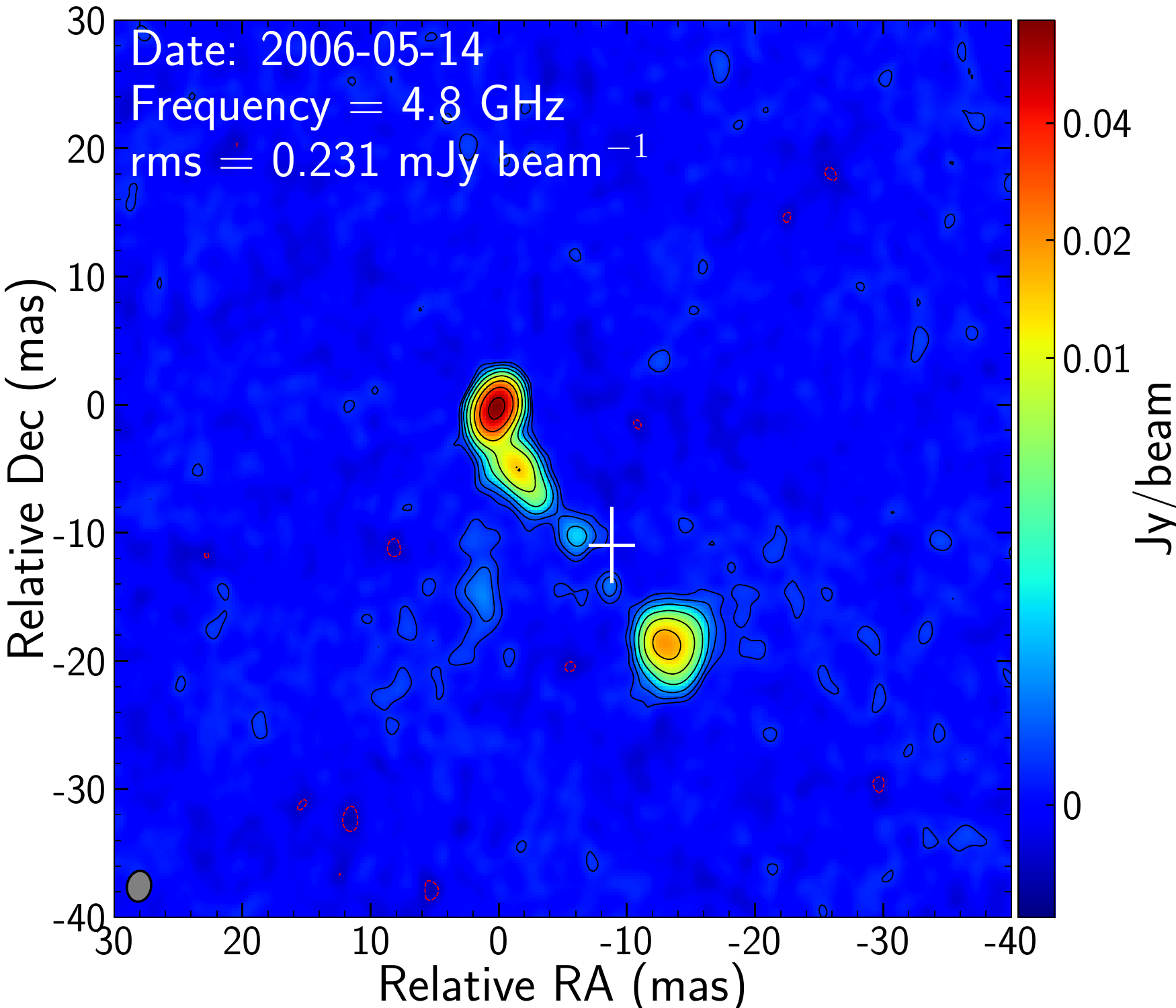} \hspace{5mm}
\includegraphics[width=0.45\textwidth, angle=0]{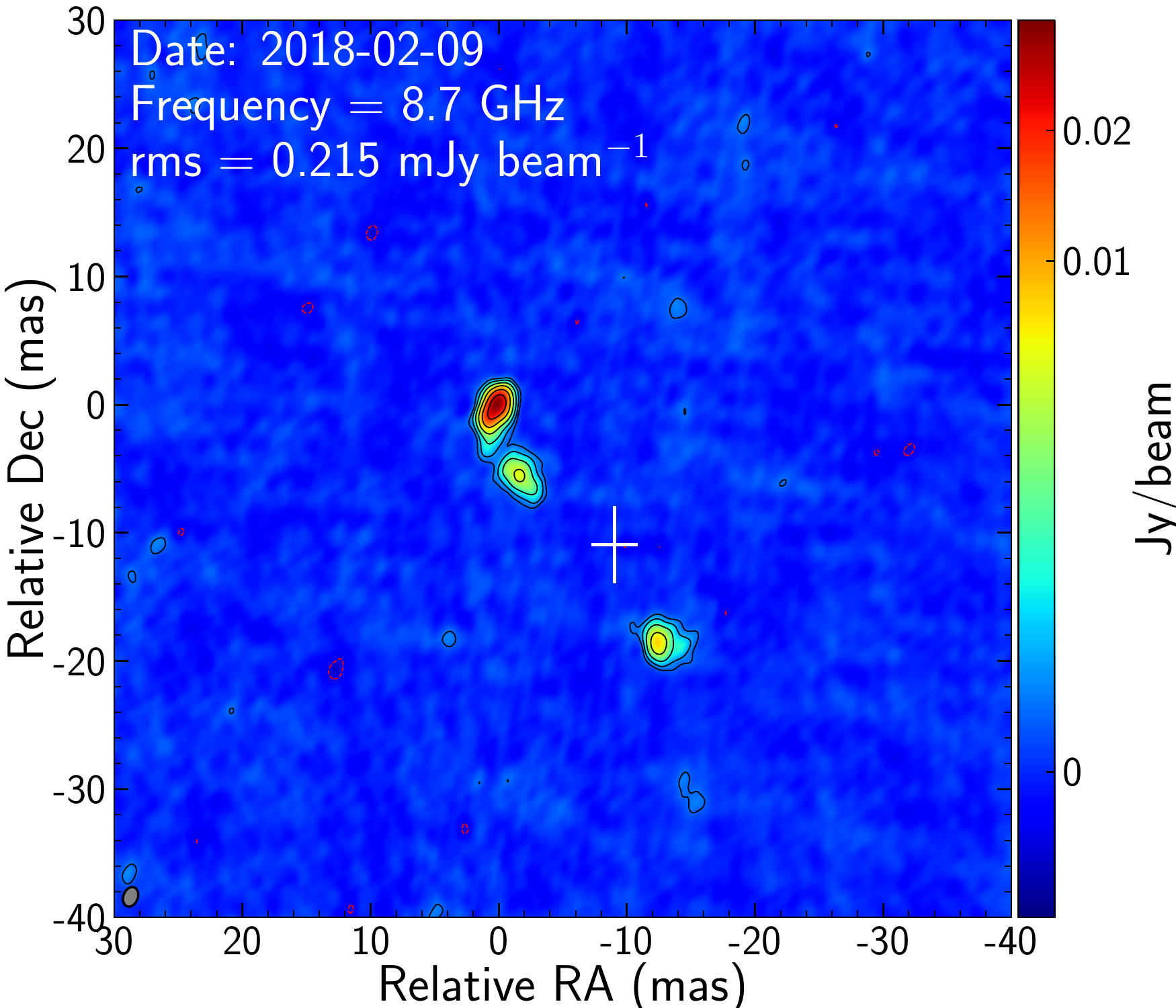}
\caption{
VLBA images of J1110$+$4817. The white cross marks the {\em Gaia} position. The size of the cross represents the uncertainties on each coordinate.  
{\it Top-left}: 1.5 GHz. 
The peak intensity is 174~mJy\,beam$^{-1}$. The lowest contours are drawn at $\pm 0.95$~mJy\,beam$^{-1}$, the positive contour levels increase by a factor of 2. The elliptical Gaussian restoring beam is 12.2~mas\,$\times$\,4.4~mas (full width at half-maximum, FWHM) at a major axis position angle $-1.1\degr$ (measured from north through east), as indicated in the bottom-left corner.
{\it Top-right}: 2.3 GHz. 
The peak intensity is 167~mJy\,beam$^{-1}$. The  restoring beam is 6.2~mas\,$\times$\,4.2~mas (FWHM) at a position angle $-22.6\degr$.
{\it Bottom-left}: 4.8 GHz. 
The peak intensity is 74~mJy\,beam$^{-1}$. The  restoring beam is 2.4~mas\,$\times$\,1.9~mas (FWHM) at a position angle $10.1\degr$.
{\it Bottom-right}: 8.7 GHz. 
The peak intensity is 36~mJy\,beam$^{-1}$. The  restoring beam is 1.6~mas\,$\times$\,1.1~mas (FWHM) at a position angle $-20.8\degr$.
}
\label{VLBIimage}
\end{figure*}

For the data calibration, we used the NRAO Astronomical Image Processing System \citep[{\sc AIPS},][]{2003ASSL..285..109G} and followed a standard procedure \citep[e.g.][]{1995ASPC...82..227D}. We calibrated the ionospheric delays and corrected for the measured Earth orientation parameters. Then we applied digital sampler corrections. We used a short 1-min data scan to determine instrumental phases and delays, and bandpass correction. Initial amplitude calibration was done using the gain curve and system temperature data from the antennas. After the final step of fringe-fitting in {\sc AIPS}, the calibrated visibility data were exported to {\sc Difmap} \citep{1997ASPC..125...77S}. Here we performed standard hybrid mapping with iterations of {\sc clean} deconvolution, phase-only self-calibration, and finally phase and amplitude self-calibration. The resulting naturally-weighted 1.5-GHz image of J1110+4817 is displayed in the top-left panel of Fig.~\ref{VLBIimage}. 

The primary goal of the VLBA project BD107 (PI: V. Dhawan) was an astrometric study of X‐ray binaries \citep{2004HEAD....8.1724D}, while the project BJ56 (PI: P. Jonker) was aimed at finding mas-scale jets in black hole X‐ray transients. Both experiments were observed very close in time, with ten antennas of the array: BD107 on 2015 January 17 and BJ56 on 2015 January 26. We downloaded the raw data from the NRAO archive and analysed them in the same way as described above.

In the first experiment, the observing frequencies were 8.4 and 15~GHz, using eight 8-MHz IFs in right circular polarization. The source J1110+4817 was observed during four short ($\sim 50$~s) scans separated by $\sim 42$~min in time. At 8.4 GHz, we could reproduce the same main features as seen in other known VLBI images of J1110+4817, but the image is less sensitive because of the limited observing resources. However, Gaussian brightness distribution model components could be fitted to the visibility data in {\sc Difmap} \citep{1997ASPC..125...77S}, allowing us to compare the component positions with those measured at the same frequency but at different epochs (see later in Sect.~\ref{morph}). On the other hand, the source was too weak at 15-GHz for successful fringe-fitting in AIPS. This is consistent with the result of \citet{2016MNRAS.459..820T} who could neither detect J1110+4817 in their 15-GHz VLBA observations. Fortunately, scans on a nearby bright and compact calibrator source J1110+4403 were also scheduled adjacent to the J1110+4817 scans, before and after them. Even though the angular separation of the two sources ($4\fdg24$) was rather large, we could attempt phase-referencing \citep{1995ASPC...82..327B}, by interpolating the delay and rate solutions obtained for J1110+4403 to the data of our object of interest, J1110+4817. The resulting phase-referenced image is severely affected by coherence losses due to the large separation of the calibrator, but we could at least determine the location of the 15-GHz brightness peak which coincides with the northeastern component of the source. This feature is also the brightest in other lower-frequency VLBI images (Fig.~\ref{VLBIimage}).

The second VLBA experiment in 2005 January (BJ56) used the frequencies 2.3 and 8.4~GHz, with $2 \times 8$-MHz IFs in dual (left and right circular) polarization. The source J1110+4403 was observed in 21 scans evenly distributed over a period of $\sim 8$~h. Each scan lasted for $\sim 50$~s. The resulting images are similar to the ones available in the Astrogeo database (see Fig.~\ref{VLBIimage}). Again, the parameters of the Gaussian model fitting to the 8.4-GHz data can contribute to the study of long-term structural changes in J1110+4403 (Sect.~\ref{morph}).

\section{Observed properties of J1110+4817}
\label{properties}

\subsection{Radio morphology and spectral index images}
\label{morph}

In the high-resolution radio structure of \tar\ (Fig.~\ref{VLBIimage}), the projected linear distance between the two dominant features, the brighter northeastern (NE) component and the weaker southeastern (SW) one, is nearly $200$~pc. In the CSO model, these would be mini-lobes of a young growing radio AGN. The brightest peaks represent the terminal hot spots where the jet heads interact with the surrounding ISM. In the 4.8-GHz image in the bottom-left panel of Fig.~\ref{VLBIimage} \citep[also seen in ][]{2007ApJ...658..203H}, an elongated feature embedded with jet knots trails back from the NE component to the geometric centre; such a morphology is very similar to the archetypal CSO OQ\,208 \citep{1997A&A...318..376S,2013A&A...550A.113W}.

\begin{figure}
    \centering
    \includegraphics[width=0.9\columnwidth]{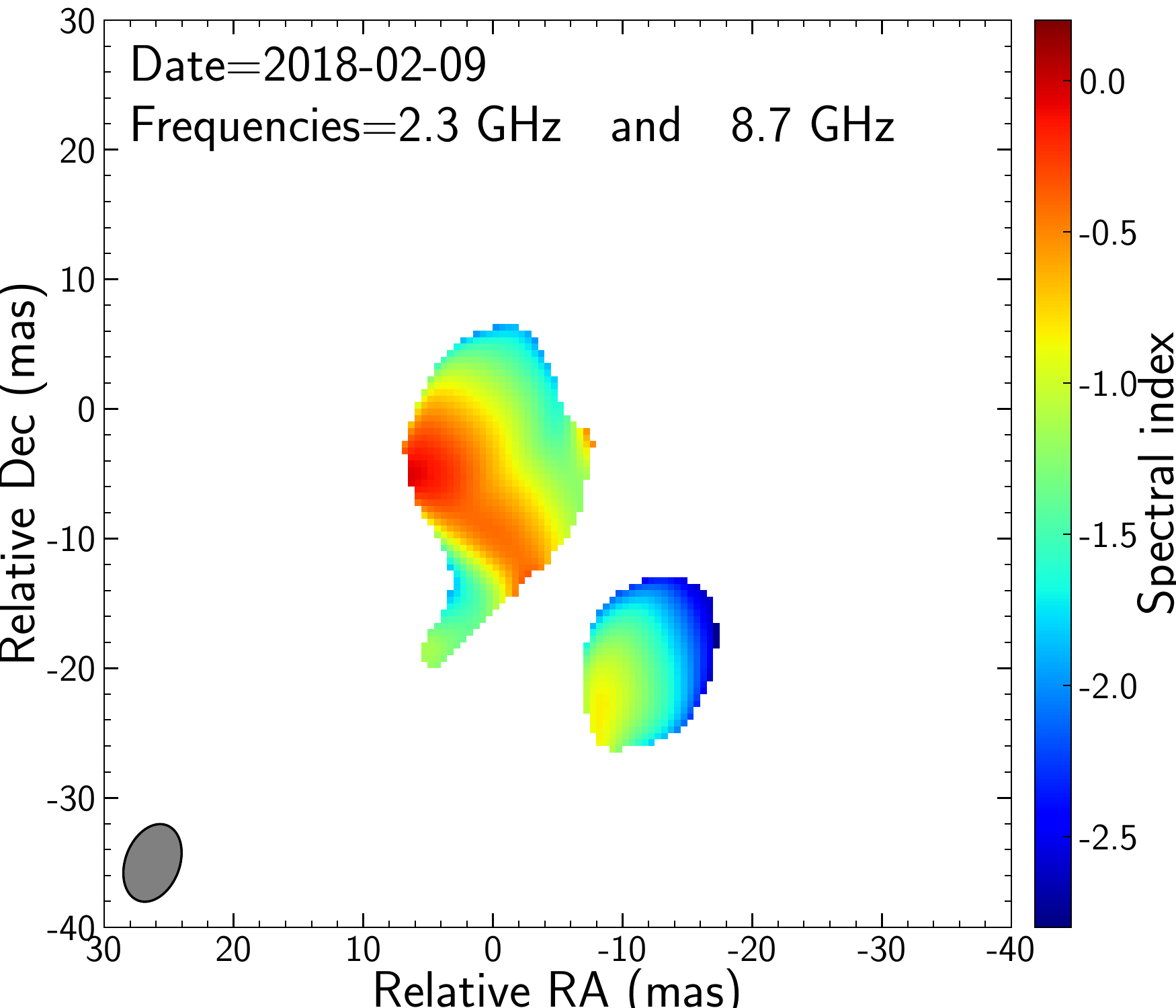}
    \caption{Spectral index map of \tar\ based on the images made from VLBA data observed at 2.3 and 8.7~GHz simultaneously.}
    \label{fig:specindex-sx}
\end{figure}

An alternative explanation would be a classical core--jet structure \citep{1979ApJ...232...34B,2019ARA&A..57..467B}, with the more compact and brighter NE feature as a synchrotron self-absorbed core and the SW feature as an optically thin jet component. VLBI cores are expected to have flat spectrum 
with spectral index $\alpha \ga -0.5$ (defined in the sense that the flux density $S$ is proportional to $\nu^{\alpha}$, with $\nu$ being the observing frequency). 
With multi-frequency VLBI images at hand, we can produce a spectral index map of \tar, as was done also by \citet{2016MNRAS.459..820T} using simultaneous observations at 4.8 and 8.4~GHz. Here we show another spectral index image, produced between 2.3 and 8.7~GHz (Fig.~\ref{fig:specindex-sx}) from simultaneous VLBA measurements made on 2018 February 9 available in the Astrogeo data base. The images at both frequencies were convolved with the same restoring beam. Our spectral index image and that of \citet{2016MNRAS.459..820T} consistently show that there is no obvious flat-spectrum core feature found in the source. While the SW component has a steeper spectrum, the generally flatter spectrum of the NE component can be explained with a hot spot emission from within the lobe.  

Probably the most intriguing feature in the lower-frequency images is the faint, steep-spectrum eastern extension apparently originating from the NE component. The emission can be traced until at least 30~mas (a projected distance of $\sim$220 pc) (Fig.~\ref{VLBIimage}) and is nearly perpendicular to the main NW--SE structure of the radio source. 
The extension is clearly seen in the 1.6- and 2.3-GHz VLBI images and is marginally detected at 4.8~GHz. It is completely resolved at 8.7~GHz, suggesting a diffuse, steep-spectrum feature (Fig.~\ref{VLBIimage}). Further illustration for the steep spectrum is the spectral index map (Fig.~\ref{fig:specindex-sx}). Such an extension is unusual but not unprecedented among CSOs.

The radio morphology we observe in \tar\ is also reminiscent of the structure in 0402$+$379, a unique pc-scale jetted binary SMBH system \citep{2006ApJ...646...49R}. It is possible to interpret the transverse structure as a steep-spectrum jet associated with a companion AGN located within the NE feature. Last but not least, the radio structure of \tar\ could result from a superposition of two unrelated AGNs seen in projection. We will consider arguments for and against all the above scenarios (i.e., CSO, core--jet structure, binary AGN, or a chance coincidence of two AGNs in the field) in Sect.~\ref{cso}, after compiling other relevant observational facts about \tar.

\subsection{Component proper motions}

\begin{figure}
    \centering
    \includegraphics[width=0.9\columnwidth]{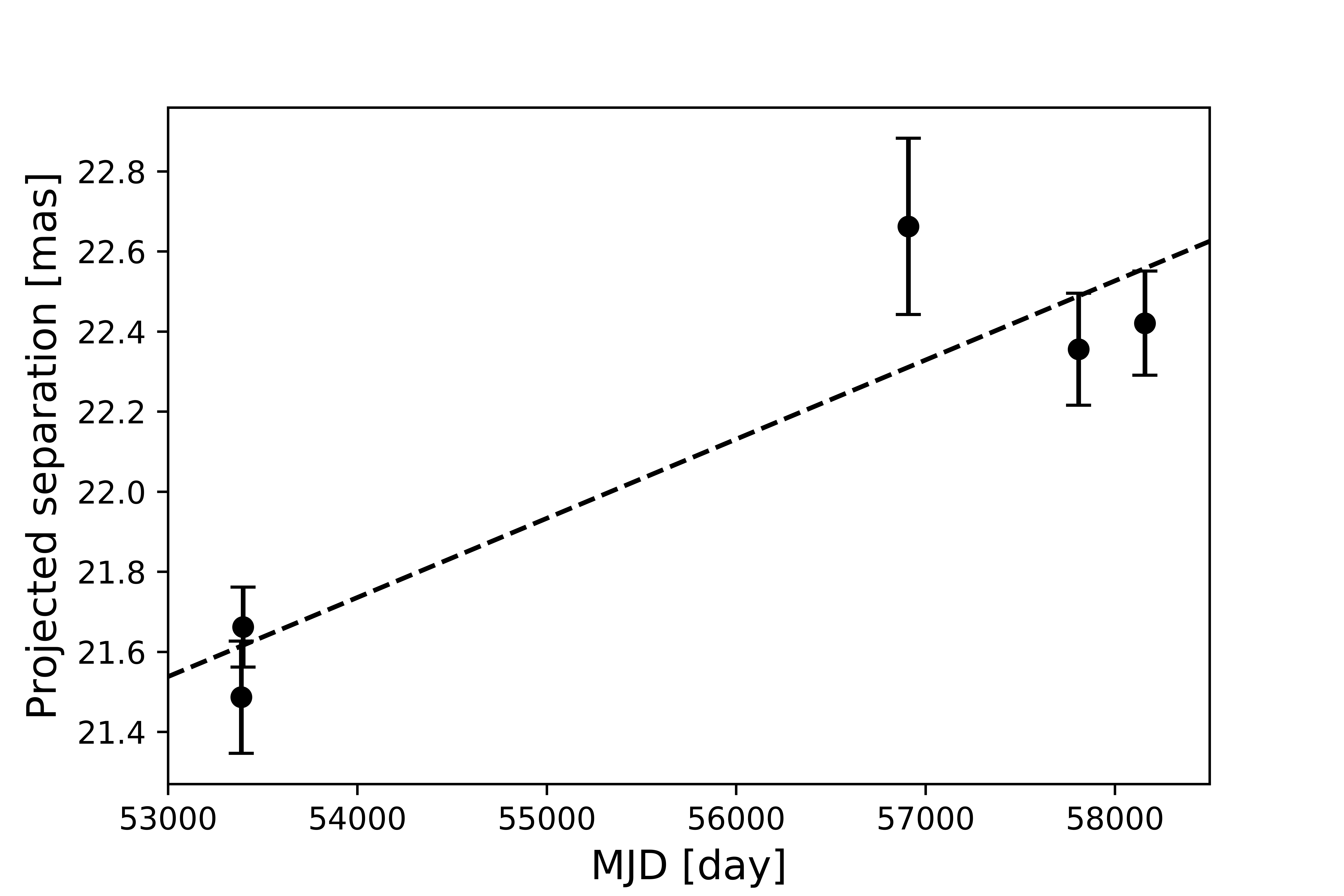}
    \caption{The projected separation between the brightest NE and the SW component as a function of time. The fitted apparent linear proper motion (dashed line) is 0.072 $\pm$ 0.019 mas\,yr$^{-1}$. }
    \label{fig:propermotion}
\end{figure}

A series of VLBI images at the 8-GHz band covering a period of $\sim 13$~yr from 2005 to 2018 allows us to reveal changes in the $\sim 10$-mas scale radio structure of \tar. Based on the relative positions of the circular Gaussian brightness distribution model components fitted to the NW and SE features in {\sc Difmap}, we list their separations in Table~\ref{tab:propermotion} and plot the values as a function of time in Fig.~\ref{fig:propermotion}. Assuming a linear change, the best-fit angular separation speed between NW and SE is $\mu = 0.072 \pm 0.019$~mas\,yr$^{-1}$. At the redshift of the source ($z=0.74$), this corresponds to an apparent superluminal motion with the speed $\beta_\mathrm{app} = 3.0 \pm 0.8$ expressed in the units of the speed of light $c$. Superluminal motion \citep{1966Natur.211..468R} is a phenomenon often observed in AGNs with relativistic jets pointing close to the line of sight  \citep[e.g.][]{1995PASP..107..803U}. Mildly superluminal hot spot advance speeds are not unprecedented among CSOs \citep[e.g.][]{2012ApJS..198....5A}.

\begin{table}
	\centering
	\caption{Angular separation between the NW and SE radio components in \tar\ from VLBI imaging and model-fitting at around 8~GHz.}
	\label{tab:propermotion}
	\begin{tabular}{lcc} 
		\hline
		Date & Frequency & Separation\\
		     & (GHz)     & (mas)\\
		\hline
		2005 Jan 17 & 8.4 & $ 21.49 \pm 0.14$ \\
		2005 Jan 26 & 8.4 & $ 21.66 \pm 0.10$ \\
		2014 Aug 09 & 8.7 & $ 22.66 \pm 0.22$ \\
		2017 Feb 24 & 8.7 & $ 22.36 \pm 0.14$ \\
		2018 Feb 09 & 8.7 & $ 22.42 \pm 0.13$ \\
		\hline
	\end{tabular}
\end{table}

\subsection{Radio continuum spectrum}

\begin{figure}
    \centering
    \includegraphics[width=0.65\columnwidth, angle=-90]{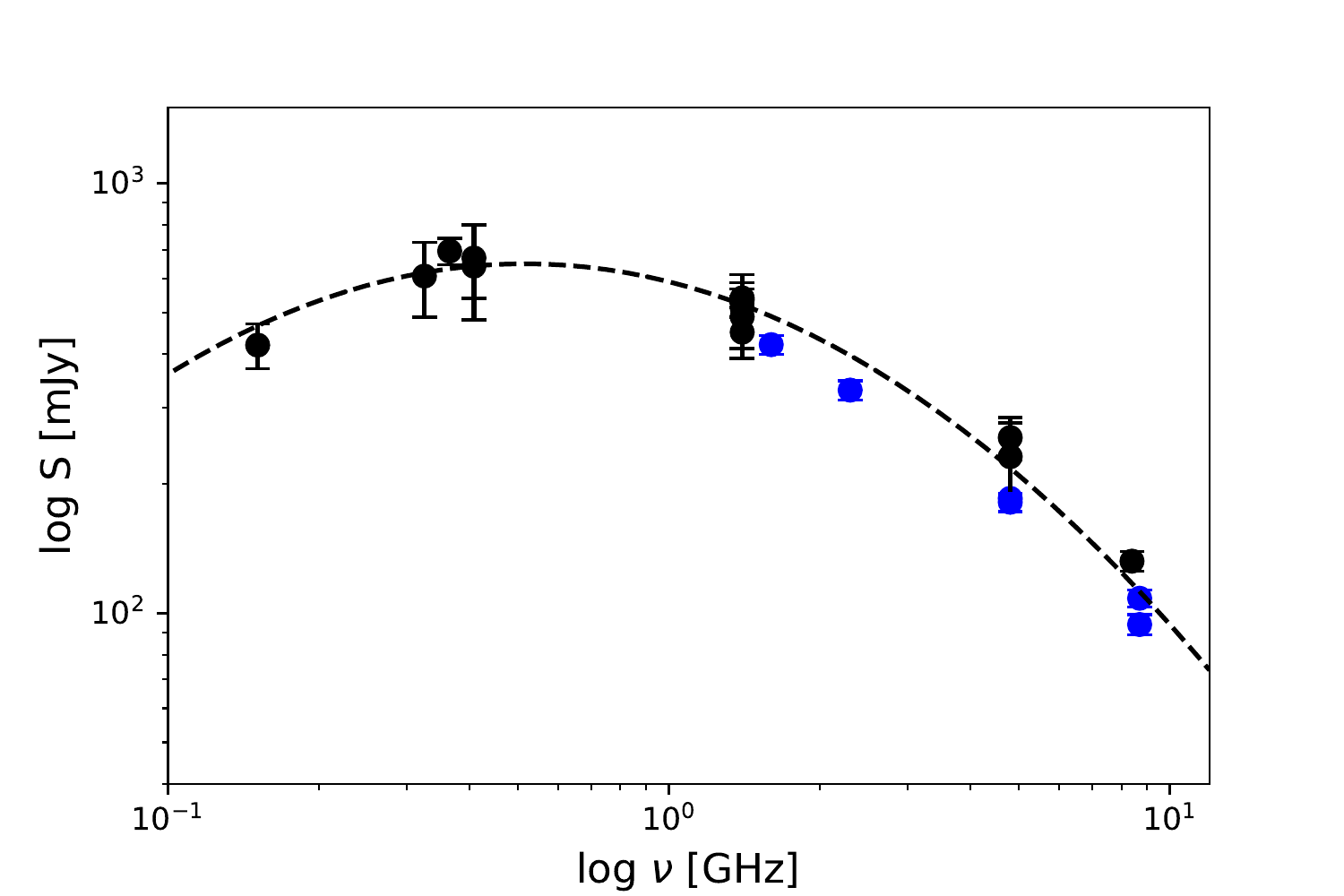}
    \caption{Radio spectrum of \tar\ peaking at $\sim 500$~MHz. The dashed curve indicates a log-parabolic fit to the total flux density data. 
    The blue-coloured symbols are VLBA flux density measurements which are lower limits to the total flux density.}
    \label{fig:spectrum}
\end{figure}

From the literature, we collected total radio flux density measurements of \tar\ (Table~\ref{tab:spectrum}) made with single-dish telescopes or connected-element interferometers at different epochs. In a few cases where the flux density errors were not available, we assumed 5 per cent uncertainties, as indicated with parentheses in Table~\ref{tab:spectrum}. 
We examined the flux densities measured at the same or similar frequencies and found no major year-scale variability. 
The broad-band radio spectrum is plotted in Fig.~\ref{fig:spectrum}. The overall shape of the spectrum is convex, and according to a log-parabolic fit in the form $\log S = a (\log \nu - \log \nu_0)^2 + b$, where $a$ and $b$ are constants \citep[e.g.][]{2017MNRAS.467.2039C}, the peak frequency is $\nu_0 \approx 500$~MHz and the peak flux density $S_0 \approx 650$~mJy. Integrated flux density values obtained with high-resolution VLBI observations are also indicated in Fig.~\ref{fig:spectrum}. These are naturally below the spectral curve derived from total flux density measurements because some extended radio emission is resolved out (undetected) on longer VLBI baselines. 

\begin{table}
	\centering
	\caption{Total radio flux densities {\em (top)} and integrated VLBI flux densities {\em (bottom)} of \tar\,. Values in parentheses indicate 5 per cent flux density errors where the measured values were not reported in the original publication.}
	\label{tab:spectrum}
	\begin{tabular}{ccl} 
		\hline
		Frequency & Flux density & Reference\\
		 (GHz)    & (mJy)        & \\
		\hline
		0.151 & $420 \pm 50$    & \citet{1988MNRAS.234..919H} \\
		0.325 & $608 \pm 120$   & \citet{1997AAS..124..259R} \\ 
		0.365 & $695 \pm 49$    & \citet{2007ApJ...658..203H} \\
		0.408 & $641 \pm 160$   & \citet{1968MNRAS.139..529P} \\
		0.408 & $670 \pm 28$    & \citet{1995MNRAS.274..324G}  \\
	    1.4   & $541 \pm (27)$  & \citet{1995ApJ...450..559B} \\ 
	    1.4   & $535 \pm 10$    & \citet{1998AAS..128..153M} \\ 
		1.4   & $513 \pm 15$    & \citet{1998AJ....115.1693C} \\
		1.4   & $489 \pm 98$    & \citet{1992ApJS...79..331W} \\
		1.4   & $450 \pm 38$    & \citet{1972AcA....22..227M} \\
		4.8   & $256 \pm 29$    & \citet{1991ApJS...75.1011G} \\
		4.8   & $231 \pm 46$    & \citet{1996ApJS..103..427G} \\
		8.4   & $132 \pm (7)$   & \citet{1992MNRAS.254..655P} \\
		\hline
		1.5   &  $421 \pm 21$   & this paper  \\
		2.3   &  $330 \pm (17)$ & \citet{2011AJ....142...89P}  \\
		4.8   &  $184.8 \pm 0.3$& \citet{2007ApJ...658..203H}  \\
		4.8   &  $181 \pm (9)$  & \citet{2011AJ....142...89P}  \\
		8.7   &  $94 \pm (5)$   & \citet{2011AJ....142...89P}  \\
		8.7   &  $108 \pm (5)$  & \citet{2000ApJ...534...90P}  \\
		\hline
	\end{tabular}
\end{table}
	
\subsection{Accurate radio and optical positions}

The current 3rd realization of the VLBI-based International Celestial Reference Frame (ICRF3, P. Charlot et al., in prep.) lists\footnote{\url{http://hpiers.obspm.fr/icrs-pc/newwww/icrf/icrf3sx.txt}} J1110+4817 with right ascension $\alpha_\mathrm{VLBI} = 11^\mathrm{h} 10^\mathrm{m} 36\fs32410 \pm 0\fs00003$ and declination $\delta_\mathrm{VLBI} = 48\degr 17\arcmin 52\farcs4498 \pm 0\farcs0003$. This position refers to the brightness peak that coincides with the NE component in the VLBI images. On the other hand, accurate equatorial coordinates of the object in the optical are found in the 2nd Data Release \citep[DR2,][]{2018A&A...616A...1G} of the
{\em Gaia} space astrometry mission \citep{2016A&A...595A...1G}. These are $\alpha_\mathrm{Gaia} = 11^\mathrm{h} 10^\mathrm{m} 36\fs32327$ and declination $\delta_\mathrm{Gaia} = 48\degr 17\arcmin 52\farcs4382$. The formal uncertainties are 1.8~mas and 3.0~mas in right ascension and declination, respectively\footnote{\url{https://gea.esac.esa.int/archive/}}. The {\em Gaia} optical position is marked in the VLBI images (Fig.~\ref{VLBIimage}) with a cross. Taking also the uncertainty of the ICRF3 VLBI position (0.3~mas in both coordinates) into account, the radio--optical offset $\Delta = 14.2$~mas is highly significant. This corresponds to $\sim 100$~pc projected linear offset. 

The optical AGN emission is known to be dominantly originating from the accretion disk. While typically sub-mas VLBI--{\em Gaia} offsets indicate the importance of pc-scale optical jet emission \citep[e.g.][]{2017A&A...598L...1K,2019ApJ...871..143P}, the large offset measured in J1110+4817 clearly suggests that an accreting supermassive black hole is located at about halfway between the NE and SW radio features seen in the VLBI images. 

\section{On the nature of J1110+4817}
\label{cso}

Based solely on the radio morphology of \tar\ revealed by multi-frequency VLBI imaging (Fig.~\ref{VLBIimage}), there are alternative scenarios to explain the physical nature of the source. Here we discuss them one by one and conclude that the CSO interpretation is the most plausible.

\subsection{Core--jet structure}

Both the size and the appearance of the VLBI structure of \tar\ would be consistent with a typical core--jet source. In this case, the brighter NE component with a flatter radio spectrum would serve as the synchrotron self-absorbed core, and the SW component would be a bright, optically thin component in the jet. The latter is apparently connected to the core with a series of weaker features most visible in the 4.8- and 8.7-GHz images (Fig.~\ref{VLBIimage}). The mildly superluminal motion, the overall peaked spectrum, and the large fraction of the total flux density recovered by VLBI could all be reconciled with the properties of a jetted quasar, although the steep-spectrum jet-like eastern extension seen on the same side of the core but at a large angle from the other jet in the lower-frequency images would require special explanation. However, the large offset between the radio and optical positions of the AGN is a strong argument against the core--jet interpretation.

\subsection{Binary AGN}

Known binary AGNs with separations $\la 200$~pc are very rare. In a study of archival VLBI data available for more than 3000 luminous radio AGNs at multiple frequencies, \citet{2011MNRAS.410.2113B} found only one case, the already known pc-scale binary 0402$+$379 with 7.3~pc separation \citep{2006ApJ...646...49R}. This object is, in fact, the only securely confirmed binary AGN resolved with VLBI to date, where \mbox{H\,{\sc i}} absorption is also mapped at high resolution \citep{2009ApJ...697...37R}, and possibly the orbital motion is detected \citep{2017ApJ...843...14B}.
An important  difference between \tar\ and 0402$+$379 is that our target lacks a secondary compact flat-spectrum core. This may be considered an argument against the radio-emitting SMBHB scenario, although it does not fully exclude it. For example, the flatter-spectrum NE feature and the steep-spectrum eastern extension could belong to a core--jet source, while part of the NE emission and the SE feature are the symmetric lobes of a CSO where the core is undetected with VLBI but coincident with the {\em Gaia} optical position. The existence of such a radio core could be tested with higher-sensitivity VLBI observations at around 5~GHz where the putative jet is also visible (Fig.~\ref{VLBIimage}). Additionally, the inclusion of shorter interferometer baselines would help to detect any potential low surface brightness feature in continuation of the eastern extension at $\sim 100$~mas angular scale.

\subsection{AGN superposition}

Similar to the binary scenario in terms of the number of separate AGNs making up the complex is if we see a superposition of two directionally very close but otherwise physically unrelated objects in \tar. In this case, they may be at significantly different redshifts, which is apparently not reflected in the optical spectrum. Moreover, a chance directional coincidence of two compact radio AGNs within $\sim 25$~mas angular separation has an extremely low probability. To estimate this, we take $N \approx 10^{4}$ as the total number of VLBI-detected extragalactic sources in the northern sky\footnote{\url{http://astrogeo.org/vlbi/solutions/rfc\_2020a/}}, corresponding to a solid angle of $\Omega=2\pi$~sr. (We exclude the southern sky from the calculation because its coverage is less complete due to the sparse VLBI networks in the southern hemisphere.) The solid angle subtended by a 25-mas radius circle around a given object is $\omega \approx 5 \times 10^{-14}$~sr. The probability of another compact radio AGN falling into this solid angle is $q = \omega / \Omega \approx 7 \times 10^{-15}$. With the given number of possible other AGNs ($N$), the probability of a chance coincidence is $p = 1 - (1-q)^N \approx 7 \times 10^{-11}$. Therefore we believe it would require some strong independent evidence before \tar\ can be seriously considered as a chance superposition of two AGNs.

\subsection{A single compact symmetric object}

The projected linear size, the morphology (Fig.~\ref{VLBIimage}), and the spectral index distribution (Fig.~\ref{fig:specindex-sx}) are generally consistent with the CSO classification of \tar. The shape and the peak frequency of the overall radio spectrum (Fig.~\ref{fig:spectrum}) are similar to typical CSO spectra \citep[e.g.][]{2012ApJS..198....5A,2016MNRAS.459..820T}. The difference between the total flux density values and the integrated VLBI flux densities indicates that the dominant fraction ($\sim 70-80$ per cent) of the emission is recovered in the components detected with high-resolution VLBI observations, also usual for other CSOs \citep{2016MNRAS.459..820T}. Although the radio core at the {\em Gaia} position is not revealed in the VLBI images presented here, because {\em Gaia} identifies the location of the accreting AGN, the symmetric NE and SW components must be lobes associated with two-sided jets in the CSO scenario. This astrometric evidence is not the only one but probably the strongest for \tar\ being a CSO. For the rest of the paper, we assume the CSO nature of the source and discuss its properties in the context of this interpretation.

\section{J1110+4817 as a CSO}
\label{discussion}

\subsection{Kinematic age}

The observed expansion of the radio structure in \tar\ (Fig.~\ref{fig:propermotion}) makes it possible to estimate the kinematic age of the source by assuming a constant speed and extrapolating the motion back in time until the two components coincide. Considering the current angular separation of $\sim 22$~mas and its rate $\mu = 0.072$~mas\,yr$^{-1}$, the kinematic age of the radio source in its rest frame is $\sim 530$~yr, a typical value for CSOs.

\subsection{The inclination of the jet}

Assuming the CSO scenario and that the AGN is located in between the two radio hot spots at the {\em Gaia} optical position, we can estimate the inclination angle of the jet axis with respect to the line of sight ($\theta$). In a simple model, the jets are moving in exactly opposite directions, the activity started simultaneously, and the plasma propagated with the same speed through the surrounding medium on both sides of the central black hole. Then, following e.g. \citet{1997ApJ...485L...9T}, the ratio of the projected distances from the core to the hot spots are
\begin{equation}
\frac{d_\mathrm{NE}}{d_\mathrm{SW}} = \frac{1 + \beta \cos \theta}{1 - \beta \cos \theta}.
\end{equation}
Here $\beta$ is the bulk speed of the jet, expressed in the units of $c$. Substituting the core--hot spot distances on the advancing and receding sides, $d_\mathrm{NE} = 14 \pm 2$~mas and $d_\mathrm{SW} = 9 \pm 2$~mas, respectively, we get
\begin{equation}
\label{eq:betacostheta}
\beta \cos \theta = \dfrac{d_\mathrm{NE}-d_\mathrm{SW}}{d_\mathrm{NE}+d_\mathrm{SW}} = 0.217 \pm 0.021.
\end{equation}
For the measurements, we used the 4.8-GHz VLBA image \citep[Fig.~\ref{VLBIimage},][]{2007ApJ...658..203H} obtained on 2006 May 14. The error bars are dominated by the uncertainty in the {\em Gaia} DR2 position determination. Since $\beta < 1$, we get $\theta \ga 77\degr$.

We measured the apparent hot spot separation speed $\beta_\mathrm{app} = 3.0 \pm 0.8$ between the NW and SE components. Substituting half of this value as the apparent jet speed on both sides of the centre into eq.~B1 of \citet{1993ApJ...407...65G}, 
\begin{equation}
\frac{1}{2} \beta_\mathrm{app} = \frac{\beta \sin \theta}{1 - \beta \cos \theta},
\end{equation}
and assuming $\beta \approx 1$, we estimate the inclination angle $\theta \approx 67\degr$. This is roughly consistent with the inference from the arm length ratio, both indicating a jet pointing away from the line of sight.

\subsection{Optical positions relative to hot spots in other CSOs}

A key argument for the CSO identification of \tar\ is the {\em Gaia} DR2 optical position of the AGN falling in the centre of the symmetric radio structure. The optical emission from the accretion disk might be obscured in many CSOs due to orientation effects and the dense surrounding medium. However, some of them can be sufficiently bright to be found in the {\em Gaia} DR2 astrometric catalogue. This allows for checking their radio and optical coordinates measured with comparable (mas or sub-mas) accuracy, to look for cases similar to \tar\ where the significant positional offset supports the CSO scenario.

For example, \citet{2016MNRAS.459..820T} presented 24 confirmed CSOs of which 15 were newly identified objects. In addition, they listed 33 unconfirmed and 52 refuted cases (our target source, \tar\ was included in the former category). Only 4 out of the 24 confirmed CSOs, and 7 out of the 85 unconfirmed/refuted sources have a {\em Gaia} DR2 catalogue entry \citep{2018A&A...616A...1G}. We found two confirmed (J0943$+$1702 and J1326$+$3154) and one additional unconfirmed (J0832$+$1832) CSOs where the {\em Gaia} position lies between the two radio hot spots. As the optical position can be related to the accretion disk around the central supermassive black hole, this provides further evidence  for the CSO nature of J0943$+$1702 and J1326$+$3154, and strongly suggests that  J0832$+$1832 is, in fact, a CSO.

One of the brightest CSOs in the sample of \citet{2016MNRAS.459..820T}, J1326$+$3154 is very similar to \tar\ in terms of radio morphology, as it shows a symmetric linear structure with three main features. The central component is a compact flat-spectrum core which also coincides with the optical position. Moreover, the extended emission of an edge-brightened lobe continues in the direction almost perpendicular to the main structural axis. In the cases of the other CSOs with {\em Gaia} coordinates available, the optical source mostly seems to be in the extension of the multi-component radio jet structure, while in a few cases it is $30-50$~mas far away from the VLBI core. These sources may be subjects of detailed follow-up studies that are beyond the scope of this paper.

\subsection{Eastern extension}

One of the most unusual features of \tar\ is the outflow-like extension to the south and southeast of the brighter lobe (see top panels of Fig.~\ref{VLBIimage}) apparently arising from the NE hot spot, in a direction nearly perpendicular to the jet. The outflow appears to have a direct connection with the NE lobe, but is not connected with  either the nucleus (i.e. the optical core position) or the SW lobe. It may be driven by the advancing jet associated with the northern lobe.

Such an extension is unusual but not unprecedented among CSOs. For example, the quasar PKS 1155$+$251 \citep{2017MNRAS.471.1873Y} shows a similar feature, where the steep-spectrum trail emission extends to a distance of $\sim 90$~pc. A possible explanation is the brightening of the hot spot in a transverse direction within the lobe due to pressure differences at the location of the jet--ISM interaction. A similar physical model was proposed by \citet{2008ApJ...684..153T} for the apparently shrinking CSO J1158+2450. Alternatively, the jet structure may have rotational motion. The bright radio AGN J2020+6163 \citep[originally classified as a core--jet source by][]{1994ApJ...425..568C} also shows a weak tail of radio emission at 1.7~GHz and, to some extent, at 5~GHz as well. That feature, almost faded away from 1982 to 1997, is connected to one of the main symmetric components of J2020+6163 that is reclassified as a CSO by \citet{2000A&A...360..887T}.

Another example for the transverse structure is the powerful radio galaxy 3C~316 \citep{2013MNRAS.433.1161A} in which an outflow of radio continuum emission arises from a lobe of the compact steep-spectrum galaxy. That feature was interpreted by radio jet-driven outflow.

\section{Summary}
\label{summary}

We studied the quasar \tar\ that was considered a CSO candidate in the literature \citep{2016MNRAS.459..820T}. Based on the analysis of archival multi-frequency VLBI imaging observations, the radio spectrum spanning nearly two orders of magnitude in frequency and peaking at $\sim 500$~MHz, as well as the location of the optical AGN emission accurately pinpointed by the {\em Gaia} space astrometry mission \citep{2016A&A...595A...1G}, we find the CSO nature of the object the most plausible scenario, although excluding the possibility of a dual radio AGN system would require further observations. Accepting the CSO interpretation of the radio source, we measure its apparent hot spot separation speed $\sim 3c$ and estimate its kinematic age $\sim 530$~yr by simply assuming a linear growth rate for the observed symmetric radio structure.

According to the lower-frequency VLBI images, there is an extended ($\sim 220$~pc) steep-spectrum radio emission in \tar\ seen nearly perpendicular to the jet axis as projected onto the sky. It appears connected to the NE lobe complex. While this type of structure is not unprecedented among CSOs, it is not widespread either. Possible explanations include an outflow driven by the expanding jet, a transverse motion of a hot spot emission caused by pressure differences in the ambient medium, the rotation of the entire CSO structure, or even the presence of a companion jetted AGN embedded or seen towards the NE lobe.

As we demonstrated in the case of \tar, the availability of accurate {\em Gaia} positional measurements opens a promising new avenue in confirming CSO candidates and studying CSOs, especially those with weak radio cores undetected in VLBI images. Even though CSOs are typically faint objects in the optical, there may be several further cases that warrant a thorough investigation of the high-resolution VLBI structure together with precise radio and optical astrometric information.

\section*{Data availability}

The datasets underlying this article were derived from sources in the public domain as given in the respective footnotes.

\section*{Acknowledgements}

We thank the anonymous referee for the timely and thought-provoking review which substantially improved the discussion of our results. This work is supported by the National Key R\&D Programme of China (2018YFA0404603), the Chinese Academy of Sciences (CAS, 114231KYSB20170003), and the Hungarian National Research, Development and Innovation Office (grant 2018-2.1.14-T\'ET-CN-2018-00001). The authors acknowledge the use of archival calibrated VLBI data from the Astrogeo Center database maintained by Leonid Petrov. The National Radio Astronomy Observatory is a facility of the National Science Foundation operated under cooperative agreement by Associated Universities, Inc. This research has made use of the NASA/IPAC Extragalactic Database (NED) which is operated by the Jet Propulsion Laboratory, California Institute of Technology, under contract with the National Aeronautics and Space Administration. 



\bibliographystyle{mnras}
\bibliography{J1110+4817}


\bsp	
\label{lastpage}
\end{document}